\documentclass[adraft]{eptcs}
\providecommand{\publicationstatus}{}
\advance\textheight 13.6pt
\advance\textheight 13.6pt
\usepackage{breakurl}             
\usepackage{underscore}           
\usepackage{wrapfig}
\newcommand{\sub}[2]{\mathclose{\left\{^{#2}\!/\!_{#1}\right\}}}
\newtheorem{exam}{Example}
\newtheorem{obse}{Observation}

\newenvironment{example}{\begin{exam} \rm }{\end{exam}}
\newenvironment{observation}{\begin{obse}\rm }{\end{obse}}

\def\titlerunning{Coinductive Validity}
\def\authorrunning{R.J. van Glabbeek}
\title{\titlerunning}
\author{Rob van Glabbeek
\institute{Data61, CSIRO, Sydney, Australia}
\institute{School of Computer Science and Engineering,
University of New South Wales, Sydney, Australia}
}
\begin{document}
\maketitle

\begin{abstract}
This note formally defines the concept of coinductive validity of judgements, and contrasts it with inductive validity.
For both notions it shows how a judgement is valid iff it has a formal proof.
Finally, it defines and illustrates the notion of a proof by coinduction.
\end{abstract}
\vspace{3ex}

\noindent
Induction and coinduction are techniques for defining sets, namely as least or greatest solutions of
certain collections of inequations, and for proving membership of thusly defined sets.
Here I deal with the special case of sets of \emph{judgements}, and inequations in the shape of \emph{proof rules}.
A \emph{judgement} can for instance be a formula in some logic, or
a statement $P \models \chi$ saying that a system $P$ satisfies a temporal formula $\chi$.
What we need to know about judgements is that they evaluate to \emph{true} or \emph{false},
and that (co)induction can be used to define the set of \emph{valid} ones---those that evaluate to \emph{true}.

I refer to \cite{San11} for a general introduction to coinduction, to \cite{JR97} for a coalgebraic
treatment of coinduction, and to \cite{KS17} for a practical introduction to the use of coinduction.
The main contribution of the present paper is a concise definition of what coinductive validity is.
Naturally, this is contrasted with inductive validity. I also define and illustrate the concept of a
coinductive proof of validity.

\paragraph{Proof rules}
Given a set $\Phi$ of potential \emph{judgements} $\psi$,
the \emph{valid} judgements are defined either inductively or coinductively by means of
a set of \emph{proof rules} of the form $\frac{\psi_1 \ \psi_2 \ \dots \ \psi_n}{\psi}$.
The guideline is:
\begin{equation}\label{validity}
\mbox{\begin{minipage}{3.1in}
A judgement $\psi$ is valid if and only if there is a rule with conclusion $\psi$ of which all premises are valid.
\end{minipage}}
\end{equation}

\paragraph{Proof}
A \emph{formal proof} of a judgement $\varphi$ is an upwardly branching tree of which the nodes are
labelled by judgements, such that the root is labelled by $\varphi$,
 and whenever $\psi$ is the label of a note $s$ and $K$ is the set of labels of the nodes directly above $s$, then
 $\frac{K}{\psi}$ is one of the given proof rules. Such a proof is \emph{well-founded} if there is no
 infinite path that keeps going up.

\paragraph{Inductive validity}
Let $\mathcal{S}_{\it ind}$ be the collection of all notions of validity $\mathcal{V} \subseteq \Phi$
that satisfy the ``if''-direction of (\ref{validity}). This collection is closed under intersections, and
hence has a least element $\mathcal{V}_{\it ind} \subseteq \Phi$. By definition, this is the notion of a valid judgement that is
inductively defined by the given rules.

\begin{observation}
A judgement is inductively valid iff it has a well-founded formal proof.
\end{observation}
It follows that the inductively valid judgements satisfy all of (\ref{validity}), and not just the ``if''-direction.

\paragraph{Coinductive validity}
Let $\mathcal{S}_{\it coind}$ be the collection of all notions of validity $\mathcal{V} \mathbin\subseteq \Phi$
that satisfy the ``only if''-direction of (\ref{validity}). This collection is closed under arbitrary unions, so
has a largest element $\mathcal{V}_{\it coind} \mathbin\subseteq \Phi$.  By definition, this is the notion of a valid judgement that is
coinductively defined by the given rules.

\begin{observation}
A judgement is coinductively valid iff it has a formal proof.
\end{observation}
It follows that the coinductively valid judgements satisfy all of (\ref{validity}), and not just the
``only if''-direction.
\bigskip

\noindent
The key difference between inductive and coinductive validity is well expressed in \cite{KS17}:\vspace{5pt}
\begin{quote}\it
  A property holds by induction if there is good reason for it to hold;\\
  whereas a property holds by coinduction if there is no good reason for it not to hold.
\end{quote}

\paragraph{Proofs by coinduction}
A method to show that a set $S\subseteq\Phi$ of judgements is coinductively valid, is by means of a
\emph{coinductive proof}: it is sufficient to construct, for each judgement $\varphi\in S$,
a nonempty proof fragment that may use all elements of
$S$ as \emph{coinduction hypotheses}, and derives $\varphi$ from them.

\begin{example}
Consider a simple process algebra with expressions $E$ given by the BNF-like grammar
\[E ::= \sum_{i\in I} a_i.E_i \mid \mu X. \sum_{i\in I} a_i.E_i \mid X \]
where $I$ is a finite index set, actions $a_i$ are drawn from an alphabet $\mathcal{A}$, and
variables $X$ from a given set {\it Var}. When $I=\{i_0\}$ is a singleton, the expression
$\sum_{i\in I} a_i.E_i$ may be abbreviated as $a_{i_0}.E_{i_0}$.

This language can be seen as a fragment of either CCS \cite{Mi90ccs}
or CSP~\cite{BHR84}. I didn't include a clause $\mu X. E$ in order to restrict to \emph{guarded recursion} \cite{Mi90ccs}.
A \emph{process} or closed expression is one in which each variable $X$ occurs within the scope of
an expression $\mu X.E$. Now let $\equiv$ be the binary relation between processes coinductively
defined by the following proof rules:
\[\frac{P_i \equiv Q_i \mbox{~for each~} i \in I}{\displaystyle\sum_{i\in I} a_i.P_i \equiv \sum_{i\in I} a_i.Q_i}
  ~~(\mbox{\sc act})
\qquad \frac{E\sub{X}{\mu X.E} \equiv Q}{\mu X.E \equiv Q}
  ~~(\mbox{\sc rec-l})
\qquad \frac{P \equiv E\sub{X}{\mu X.E}}{P \equiv \mu X.E}
  ~~(\mbox{\sc rec-r}).  \]
{\makeatletter
\let\par\@@par
\par\parshape0
\everypar{}
\noindent
Formally, the first rule is a \emph{schema}, with an instance for each choice of a finite index set
$I$, actions $a_i$, and processes $P_i$ and $Q_i$. Likewise, the other two rules are

\begin{wrapfigure}[12]{r}{0.414\textwidth}\footnotesize\vspace{-7ex}
\[\frac{
  \displaystyle\frac{
  \displaystyle\frac{
  \displaystyle\frac{
  \displaystyle\frac{
  \displaystyle\frac{
  \displaystyle\frac{
  \displaystyle\frac{
  \displaystyle\frac{
  \displaystyle\frac
  {\displaystyle \mu X.a.a.X \equiv \mu Y.a.a.a.Y}
  {\displaystyle a.\mu X.a.a.X \equiv a.\mu Y.a.a.a.Y}\makebox[0pt][l]{~~(\mbox{\sc act})}}
  {\displaystyle a.a.\mu X.a.a.X \equiv a.a.\mu Y.a.a.a.Y}\makebox[0pt][l]{~~(\mbox{\sc act})}}
  {\displaystyle \mu X.a.a.X \equiv a.a.\mu Y.a.a.a.Y}\makebox[0pt][l]{~~(\mbox{\sc rec-l})}}
  {\displaystyle a.\mu X.a.a.X \equiv a.a.a.\mu Y.a.a.a.Y}\makebox[0pt][l]{~~(\mbox{\sc act})}}
  {\displaystyle a.\mu X.a.a.X \equiv \mu Y.a.a.a.Y}\makebox[0pt][l]{~~(\mbox{\sc rec-r})}}
  {\displaystyle a.a.\mu X.a.a.X \equiv a.\mu Y.a.a.a.Y}\makebox[0pt][l]{~~(\mbox{\sc act})}}
  {\displaystyle \mu X.a.a.X \equiv a.\mu Y.a.a.a.Y}\makebox[0pt][l]{~~(\mbox{\sc rec-l})}}
  {\displaystyle a.\mu X.a.a.X \equiv a.a.\mu Y.a.a.a.Y}\makebox[0pt][l]{~~(\mbox{\sc act})}}
  {\displaystyle a.a.\mu X.a.a.X \equiv a.a.a.\mu Y.a.a.a.Y}\makebox[0pt][l]{~~(\mbox{\sc act})}}
  {\displaystyle\frac
  {\displaystyle a.a.\mu X.a.a.X \equiv \mu Y.a.a.a.Y \rule{0pt}{9pt}}
  {\mu X.a.a.X \equiv \mu Y.a.a.a.Y}\makebox[0pt][l]{~~(\mbox{\sc
      rec-l})}}\makebox[0pt][l]{~~(\mbox{\sc rec-r})}
  \hspace{35pt}\mbox{}\]
\end{wrapfigure}
\noindent
schemata with an instance for each variable $X$, expression $E$ and process $P$ or $Q$.
Here $E\sub{X}{\mu X.E}$ denotes the result of substituting $\mu X.E$ for $X$ in the expression $E$.

It is not hard to show that $\equiv$ coincides with the familiar notion of
\emph{strong bisimulation equivalence} \cite{Mi90ccs}.
Here I merely give proof by coinduction of the statement
$\mu X.a.a.X \equiv \mu Y.a.a.a.Y$.

With $S\subseteq\Phi$ a singleton, I merely need to give a nonempty proof fragment of this equation,
allowing itself as coinduction hypothesis. It is shown on the right.
\par}
\end{example}

\bibliographystyle{eptcsalpha}
\bibliography{coinduction}

\begin{thebibliography}{BHR84}
\providecommand{\bibitemdeclare}[2]{}
\providecommand{\surnamestart}{}
\providecommand{\surnameend}{}
\providecommand{\urlprefix}{Available at }
\providecommand{\url}[1]{\texttt{#1}}
\providecommand{\href}[2]{\texttt{#2}}
\providecommand{\urlalt}[2]{\href{#1}{#2}}
\providecommand{\doi}[1]{doi:\urlalt{http://dx.doi.org/#1}{#1}}
\providecommand{\bibinfo}[2]{#2}

\bibitemdeclare{article}{BHR84}
\bibitem[BHR84]{BHR84}
\bibinfo{author}{S.D. \surnamestart Brookes\surnameend},
  \bibinfo{author}{C.A.R. \surnamestart Hoare\surnameend} \&
  \bibinfo{author}{A.W. \surnamestart Roscoe\surnameend}
  (\bibinfo{year}{1984}): \emph{\bibinfo{title}{A theory of communicating
  sequential processes}}.
\newblock {\sl \bibinfo{journal}{Journal of the ACM}}
  \bibinfo{volume}{31}(\bibinfo{number}{3}), pp. \bibinfo{pages}{560--599},
  \doi{10.1145/828.833}.

\bibitemdeclare{article}{JR97}
\bibitem[JR97]{JR97}
\bibinfo{author}{Bart \surnamestart Jacobs\surnameend} \& \bibinfo{author}{Jan
  \surnamestart Rutten\surnameend} (\bibinfo{year}{1997}):
  \emph{\bibinfo{title}{A Tutorial of (co)Algebras and (Co)Induction}}.
\newblock {\sl \bibinfo{journal}{EATCS Bulletin}} \bibinfo{volume}{62}, pp.
  \bibinfo{pages}{1132--1152}.
\newblock \urlprefix\url{https://www.cs.ru.nl/B.Jacobs/PAPERS/JR.pdf}.

\bibitemdeclare{article}{KS17}
\bibitem[KS17]{KS17}
\bibinfo{author}{Dexter \surnamestart Kozen\surnameend} \&
  \bibinfo{author}{Alexandra \surnamestart Silva\surnameend}
  (\bibinfo{year}{2017}): \emph{\bibinfo{title}{Practical coinduction}}.
\newblock {\sl \bibinfo{journal}{Math. Struct. Comput. Sci.}}
  \bibinfo{volume}{27}(\bibinfo{number}{7}), pp. \bibinfo{pages}{1132--1152},
  \doi{10.1017/S0960129515000493}.

\bibitemdeclare{incollection}{Mi90ccs}
\bibitem[Mil90]{Mi90ccs}
\bibinfo{author}{R.~\surnamestart Milner\surnameend} (\bibinfo{year}{1990}):
  \emph{\bibinfo{title}{Operational and algebraic semantics of concurrent
  processes}}.
\newblock In \bibinfo{editor}{J.~\surnamestart van Leeuwen\surnameend}, editor:
  {\sl \bibinfo{booktitle}{Handbook of Theoretical Computer Science}},
  chapter~\bibinfo{chapter}{19}, \bibinfo{publisher}{Elsevier Science
  Publishers B.V. (North-Holland)}, pp. \bibinfo{pages}{1201--1242}.
\newblock \bibinfo{note}{Alternatively see{ \em Communication and Concurrency},
  Prentice-Hall, Englewood Cliffs, 1989, of which an earlier version appeared
  as{ \em A Calculus of Communicating Systems}, LNCS 92, Springer, 1980,
  doi:\href{http:dx.doi.org/10.1007/3-540-10235-3}{10.1007/3-540-10235-3}}.

\bibitemdeclare{book}{San11}
\bibitem[San11]{San11}
\bibinfo{author}{Davide \surnamestart Sangiorgi\surnameend}
  (\bibinfo{year}{2011}): \emph{\bibinfo{title}{Introduction to Bisimulation
  and Coinduction}}.
\newblock \bibinfo{publisher}{Cambridge University Press},
  \doi{10.1017/CBO9780511777110}.

\end{thebibliography}
\end{document}